\documentclass[proof]{WileyASNA-v1}

\articletype{Article Type}%

\received{26 April 2016}
\revised{6 June 2016}
\accepted{6 June 2016}

\begin{document}

\title{Unexpected gap creating two peaks in the periods of planets of metal-rich sunlike single stars
%    title\protect\thanks{%This is an example for title footnote.}
    }

\author[1]{Stuart F. Taylor*}
% \author[2,3]{Author Two} \author[3]{Author Three}

\authormark{S.F. TAYLOR} % \textsc{et al}}

\address[1]{\orgdiv{Participation Worldscope
  }, %\orgname{},
  \orgaddress{\state{  Cottonwood, Arizona},
  \country{USA, and Hong Kong, SAR PRC}
%  \orgaddress{\state{and Hong Kong}, \country{SAR PRC}
  }}

%\address[2]{\orgdiv{Org Division}, \orgname{Org Name}, \orgaddress{\state{State name}, \country{Country name}}}
%\address[3]{\orgdiv{Org Division}, \orgname{Org Name}, \orgaddress{\state{State name}, \country{Country name}}}

\corres{*Corresponding author: Stuart F. Taylor, \email{astrostuart@gmail.com}}

\presentaddress{Sheung Wan, Hong Kong SAR PRC}

\abstract{The pileup of planets at periods of roughly one year and beyond 
is actually a bimodal peak with a wide, sharp gap
splitting the peak of the pileup in a major population of 
large planets.
Consisting of nearly 40\% of planets with periods past 200 days,
the periods of the planets of metal-rich 
stars like the sun in surface gravity which do not have a stellar companion
show two strong peaks separated by a sparsely populated region.
Monte Carlo tests show that this structure is unlikely to occur
in random distributions, 
and a comparison with objects from all the other populations show
that this feature is unlikely to be due to observational effects.
The peaks have their highest density next to the gap.
These two peaks are most strongly seen in single-planet systems,
though the gap persists in multiple planet systems.
These features are likely characteristic of planets with masses
not too much lower than Jupiter, and perhaps not too much higher.
The presence of well-defined features in the period distribution 
show that planet formation may be much more uniform than previously
expected.
}
%----\\
%In what might be called the major subpopulation of planets, 
%the ``pileup'' of planets starting at periods under one year
%and continuing to well past periods of 1000 days 
%is actually two peaks separated by a deep gap.
%%Planets periods have a main pileup between 200 and 2000 days when counted logarithmically. 
%----\\
% Planet parameters vary depending on whether the host star has aged, or evolved, to have low surface gravity, or whether the metallicity is greater or less than solar. A comparison of the pileup in logarithmic period for planets hosted by unevolved stars with metallicity richer versus poorer than solar was performed, leading to the surprising finding that among planets of single unevolved stars with metallicity as rich or richer than the sun, the counts do not form a single continuous pileup but instead are found in a bimodal pair of pileups separated by a wide gap. These two peaks contain a large majority of middle-mass planets in single-planet systems. This prominent feature presents an unexpected challenge to planet formation theories to explain its origin.}
% ~/mg/awrite/distributions/scipap_counts/GAPREAL_NATURE.doc 

\keywords{planetary systems -- exoplanets -- planet parameter distributions} %, keyword3, keyword4}

\jnlcitation{\cname{%
\author{Taylor S.F.} %, and 
%{Williams K.}, 
%\author{B. Hoskins}, 
%\author{R. Lee},  and 
% \author{G. Masato},
%\author{T. Woollings}
  } (\cyear{2019}), 
\ctitle{Unexpected gap creating two peaks in the periods of planets of metal-rich sunlike single stars}, \cjournal{A.N.}, \cvol{2018;00:1--x}.}
%\ctitle{A regime analysis of Atlantic winter jet variability applied to evaluate HadGEM3-GC2}, \cjournal{Q.J.R. Meteorol. Soc.}, \cvol{2017;00:1--6}.}

%%\fundingInfo{Funding info text.}

\maketitle

% \footnotetext{\textbf{Abbreviations:} ANA, anti-nuclear antibodies; APC, antigen-presenting cells; IRF, interferon regulatory factor}

\section{Introduction}\label{sec1}

%1~\cite{gon97,sant03,Fis05,udr03} ,3~\cite{Paivio1975}\\
%text~\citet{gon97,sant03,Fis05,udr03}\\
%parenthesis~\citep{gon97,sant03,Fis05,udr03}
%parenthesis~\citep{tay12b,tay13b}

Planets form beyond a distance where materials can condense, so it is expected 
in the counts of exoplanets by log period
there will be a main pileup beyond a range of low counts.
The observed pileup in the log period counts of exoplanets at periods longer than a few hundred days is expected,
but we report that for the planets of single stars like the sun in surface gravity
and have similar or higher
metallcity than the sun 
(``rSLSS'' planets),
this pileup is in fact split into two pileups in log period,
with a sizable gap separating two peaks.
This bimodal distribution of the pileup in such a significant fraction of planets is completely unexpected.

It is not surprising that the planets of stars that are
more metal rich than the sun have a different distribution than those of stars that are relatively metal poor.
One of the earliest properties found for stars hosting giant planets was the ``planet-metallicity'' correlation, that stars with higher iron abundance are more likely to host planets~\citep{gon97,sant03,Fis05,udr03}.
The finding that planets of stars that are more metal-rich than the sun
are in higher eccentricity orbits than are those planets associated with
stars poorer in metals has been interpreted to mean that these two groups
of objects represent two separate
populations~\citep{daw13,tay12b,tay13b}.%~\citep{daw13,tay12,tay13b,tay12b}. (DM13, T12, T13.)
%%It is known that planets of stars that are more metal-rich than the sun have higher eccentricity distributions than those of less metal-rich stars~\citep{daw13,tay12b,tay13b}.%~\citep{daw13,tay12,tay13b,tay12b}. (DM13, T12, T13.)
%Text ref (Gonzalez 1997; Santos et al. 2003; Fischer \& Valenti 2005; Udry \& Santos 2007). 

  In this work we present the structure in the period distribution of
these rSLSS objects.
We will address the eccentricities of these objects
in the context of the regions presented here in the next work.

We begin, in Section~\ref{sec_Sels},
by describing the main pileups of the main populations of ``objects'',
where ``objects'' refers to the set of parameters describing 
one planet, its orbit, and host star.
The unlikelihood of the gap is explained here.
In Section~\ref{sec_Loner}
we further divide the population with the two peaks separated by a gap,
finding a subpopulation that shows the two peaks more distinctly.
The lack of a peak in the other main populations is shown in 
Section~\ref{sec_othrSngPile},
followed by demonstrations that the gap is unlikely in
Section~\ref{sec_unlikelihood}. %{ssec_Unlikely}.
We conclude by demonstrating how these features show that
planet formation produces a surprising uniformity of results in
a large population of large planets.

%First fig was here:
\begin{figure}[t]
  \includegraphics[width=0.48\textwidth]{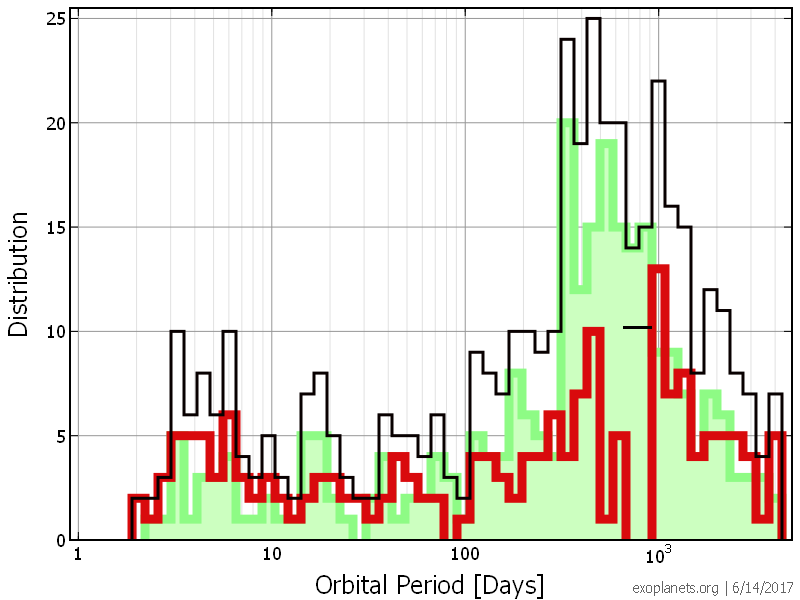} %media/ note image2<->3.png
   %21097 Aug 10  2017 ~/mg/awrite/distributions/figs/gaphst_figs/showGapLines/fig1riFesun180xcl257all435bins54_Line.png
%	\centerline{\includegraphics[width=78mm,height=9pc,draft]{media/image1.png}}%.png
	\caption{A histogram of all objects (black) shows the main pileup peaking at periods of between 300 and less than 1000 days but have a notch giving a hint of bimodal constituent. 
	Also shown are the rSLSS objects (red) which have the bimodal peak, 
	and the sum of all others (green, filled), 
                                 %(green\textbf{, filled}), %rmvBold 2019-07 
	which have a single peak. The width of the deep gap in log period space is shown as the horizontal black line. \label{fig_AllOthGap}}
\end{figure}

%Only Sunlike Objects, rSLSS&pSLSS Counts Vs Period figure; Orig 3nd fig here:
%  image3 probably ~/Dropbox/awrite/distributions/figsAvailable/gapDblPk54binsIncl2015_edtd.png
\begin{figure}[t]
  \includegraphics[width=0.48\textwidth]{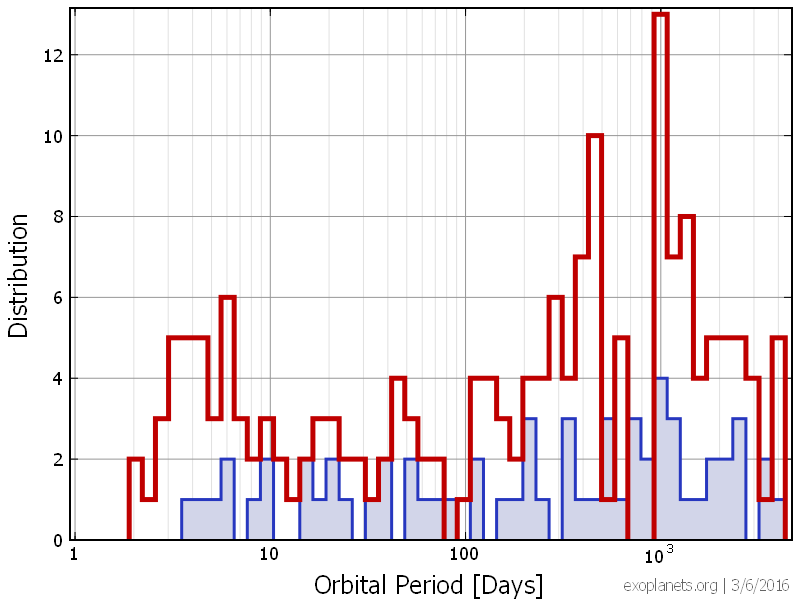}  %media/   2<->3 note no image2.png
%	\centerline{\includegraphics[width=78mm,height=9pc,draft]{media/image1.png}}%.png
  \caption{The distribution of periods of the 243 planets of single sunlike stars, with objects of 180 iron-rich stars shown in red (not filled) and planets of 63 iron-poor stars in blue (filled). The bins have been placed to fit four bins in the boundary of a gap from periods of 493.7 to 923.8 days. There is a deep gap of zero planets from 653.22 to 923.8 day periods, which is a little larger than the last two bins in the gap, that has zero planets. \label{fig_FeVsPer}}
\end{figure}

\section{Selections, one with a Gap}\label{sec_Sels}
% \section{Selections and their counts}\label{sec2}
\subsection{Discovery of a low density region}\label{ssec_discovery}
%Mvd  These patterns in the counts were found when following up on the finding that among sunlike stars the eccentricity rises with metallicity \citep{tay12b, tay13b}.
%Mvd  %(T12, T13). 
%Mvd  It was found that there was a region where there were few, and in a wide part of that region, no metal rich planets of stars without stellar companions.

% The region without objects 
The peak-gap-peak feature was discovered while investigating how long in
period the correlation of eccentricity with metallicity remained a correlation.
It had been found at short periods,
less than 100 days,~\citep{daw13,tay12b} % T12 and DM13) 
%Check how high in period T12 really went
that the eccentricities of sunlike (SL) objects
more metal rich that the sun (rSL) tended higher than the 
eccentricities of metal poor sunlike objects (pSL).
Later work~\citep{tay13b} %(T13)
found that binary stars (BS) have higher
eccentricities than single stars (SS), so~\citet{tay13b} % T13
subsequently focused on sunlike single stars (SLSS).
%The earlier works included not just the single stars (SS)
While this pattern appears to continue to periods above 100 days,
that is the means of the eccentricies of rSLSS objects are higher than those
of pSLSS objects,
it was found that at periods above 500 days there was a range where the
eccentricities of pSLSS objects peaked, having some of the highest eccentricities.
(The statistical significance of these eccentricity features is the subject
of the next paper.)
It was found, however, that in much of the period range of the
peak pSLSS eccentricities
that there are suddenly too few rSLSS objects to easily compare eccentricities.
Not only do we now find only six rSLSS objects with periods from 493.7 to 923.8 days,
we find zero rSLSS objects with periods in between the periods of 
653.8 to 923.8 days. % 653.21997
Starting at the period of 923.8 days, we find the highest density of
of rSLSS objects' periods per log period
than anywhere in the period distribution so far measured.

%When checking for how far in period the eccentricity correlation with metallicity extended, it was found that there were few values of rSL compared to pSL in the region above 500 d despite there being many rSL periods below the period of 493.7 d.
%% This extended to above 900 day, where suddenly at the period of 928.8 days there were many objects only slightly above this period.
%Above the period of 928.8 days there suddenly were many objects with periods
%only slightly longer.
% (T13, maybe T12) 
%Later, after T13, it was found that many of the few objects in the gap were BS
%objects. T13 noted that the BS objects had higher mean eccentricity. Upon removing the BS objects, there were only six rSLSS objects, which have a lower mean eccentricity than the pSLSS objects in the same region.
%There were only six objects
%We present the gap in this work, to be followed by further investigation of how the eccentricity appears to be influenced by the peak-gap-peak structure.

\subsection{Data source and selections}\label{ssec_source+sels}
We use data downloaded from exoplanets.org \citep{han14} on 2017 January 31,
which includes 434 total objects found by the 
radial velocity (RV) before 2016 at all periods up to 5000 days.
We follow the convention of these planet parameters catalogs to refer to the combined set of data for 
each planet, including data on the orbit and host star, an ``object''.
We limit our study to only those objects with periods up to 5000 days day due to there being less data on objects with longer periods. 
The binned counts of these object's periods is
histogrammed in Figure~\ref{fig_AllOthGap}, (black bins). % fig_AllOthGap=fig1

We study this ``region of interest'' (ROI) from 100 to 5000 day periods (1.7 in log period) that contains the main pileup of most of the planets, 
in which there are 313 objects.
We choose this ROI to include periods longer than the
shorter period ``valley'' region  
which has been characterized as having a paucity of objects (DM13) that goes up to 100 or 200 day periods .

We find new, unexpected features in the rSLSS selection of these objects, for planets of
metal-rich sunlike stars, or ``rSLSS'' objects,
where the ``metallicity'' of an object is that of the star.

We go through the steps we use to get to 113 rSLSS objects out of the 313 objects %explain
within the ROI that we divide from the ``other'' 198 objects hosted by stars with other parameters as explained below using Table~\ref{tab_CntDnsRgn}: %Table 1:

We note that the sensitivity of RV observations only goes down to a rough minimum
in projected planet mass, ``$m$sin$i$'', and this minimum planet mass 
rises by period.

We divide the 434 objects by whether or not their stars' surface gravity is  
equal to or above log $g$ of 4.
This yields 190 objects hosted by stars with higher surface 
gravity that we call ``sunlike'' (SL) objects,
after removing 123 low surface gravity (LSG) objects. 
We then divide these 190 SL objects into 156 single-star (SLSS) and 34 binary-star (SLBS) objects depending on whether the star has one or more stellar companions.
We further divide the 156 SLSS objects into 41 ``pSLSS'' objects where the star is metal-poor and 115 ``rSLSS'' objects where the star is metal-rich relative to the sun. Lower-case abbreviations are used to delineate selections within SLSS (or SLBS) objects. 
The distribution of counts of rSLSS and pSLSS objects are compared in Figure~\ref{fig_FeVsPer}.  %!!!
To keep the sample from having objects too far apart in effective temperature $T_{eff}$,
and to be consistent with similar work,
we apply a limit requiring objects have $T_{eff}$ in the range of 4500 to 6500 K,
which eliminates two objects  from our rSLSS selection.
We cut these even though they both have periods within each of 
the two peaks presented here, 
such that their inclusion would strengthen the case for there being two peaks. 
%standard effective stellar temperature
This produces a sample of 113 rSLSS objects in the ROI, which are shown in red in Figure~\ref{fig_AllOthGap} and subsequent histograms.
% Not Figure 3.   Not  Figure~\ref{fig_AllOthGap} 
The mean density of rSLSS objects in the ROI is then 18.1 objects per gap-width, so the count of six objects in the gap is 0.33 of the mean ROI density, which we show is significantly low. In Figure~\ref{fig_AllOthGap} we use bins of one-quarter gap-width to show how the highest densities in the ROI are right next to the gap; here, the mean density per quarter-gap-width bin is 4.5 objects.
We compare the distribution of the log periods of the rSLSS objects 
with the log periods of the 198 ``other'' objects, shown in green in Figure~\ref{fig_AllOthGap}, consisting of 41 pSLSS, 34 SLBS, and 123 LSG objects.
 %Thus, even though the 688 day period of Mars is inside the deep part of the gap, a Mars-mass planet in this period range would not be found.

%Old loc Fig (AN sbmsn) Metallicity Vs Period figure; 

%-is old as of 2019-July - Might move to when compare other pileups
The gap and two peaks can also be seen when we show the metallicity [Fe/H] versus the log period in 
Figure~\ref{fig_Fe_v_Per}. % err fig_Sunlike  %rmvBold 2019-07
%\textbf{Figure~\ref{fig_Fe_v_Per}}. % err fig_Sunlike  %rmvBold 2019-07
For perspective, we show a more full region, though this does make the gap
region appear smaller than if the figure zoomed in to the gap region.
This figure shows a ``gap'' region in period with few SLSS objects (filled circles) with [Fe/H] above the dividing line that for simplicity we set at an [Fe/H] of zero,
but the line actually would be better to be drawn at [Fe/H] of -0.03.
%(though for simplicity we use zero to define metal ``rich'' and poor objects).
Yet in this period region there is no reduction in SLSS objects with [Fe/H] below -0.03 or in LSG objects (green crosses), and only a possible small reduction in BS objects (red unfilled circles). 
% actually above [Fe/H] of -0.03)
% Jumped early peaks and gap
% Describe higher peaks here 
% Peaks here?

%First fig was here.

%First table
\begin{center}
\begin{table*}[t]%
\caption{Boundaries, counts, and densities of selections 
in the full ROI followed by the short period peak (SPP), the (shallow) gap between the peaks, and the long period peak (LPP). The counts are given, followed by the counts divided by how many gap-sized bins the range covers, which is a measure of the density of counts.\label{tab_CntDnsRgn}}
\centering
\begin{tabular*}{500pt}{@{\extracolsep\fill}lccD{.}{.}{3}ccc@{\extracolsep\fill}}
\toprule
%&\multicolumn{2}{@{}c@{}}{\textbf{Spanned heading\tnote{1}}} & \multicolumn{2}{@{}c@{}}{\textbf{Spanned heading\tnote{2}}} \\\cmidrule{2-3}\cmidrule{4-5}
%\textbf{col1 head} & \textbf{col2 head}  & \textbf{col3 head}  & \multicolumn{1}{@{}l@{}}{\textbf{col4 head}}  & \textbf{col5 head} & \textbf{col6 head} & \textbf{col7 head}  \\ %& \textbf{col6 head}  \\
%\midrule
%row no:1
 \textbf{Regions}  &  \textbf{ROI} &  \textbf{SP Tail} &  \textbf{SP Peak} &  \textbf{Gap} &  \textbf{LP Peak} &  \textbf{LP Tail}\\
%row no:2
\midrule
 Start of period, days &  100 &  100 &  263.8 &  493.7 &  923.8 &  1728.6\\
%row no:3
 End of period, days &  5000 &  263.8 &  493.7 &  923.8 &  1728.6 &  5000.0\\
%row no:4
 Width of region, in log period &  1.70 &  0.42 &  0.27 &  0.27 &  0.27 &  0.46\\
%row no:5
 Fraction of ROI width &  1.00 &  0.25 &  0.16 &  0.16 &  0.16 &  0.27\\
%row no:6
 Width relative to gap &  6.24 &  1.55 &  1.00 &  1.00 &  1.00 &  1.70\\
\midrule
 %row no:7
 $m$sin$i$ limited loner rSLSS counts &  56 &  7 &  21 &  0 &  19 &  9\\
%row no:8
 Density in Counts per gap binwidth &  9.0 &  4.5 &  21.0 &  0.0 &  19.0 &  5.3\\
%row no:9
 \  No low $m$sin$i$ loner rSLSS counts &  61 &  8 &  21 &  1 &  19 &  12\\
%row no:10
 \  Density in Counts per gap binwidth &  9.8 &  5.2 &  21.0 &  1.0 &  19.0 &  7.1\\
%row no:11
 \  No high $m$sin$i$ loner rSLSS counts &  58 &  8 &  21 &  1 &  19 &  9\\
%row no:12
 \  Density in Counts per gap binwidth &  9.3 &  5.2 &  21.0 &  1.0 &  19.0 &  5.3\\
%row no:13
 \  Multiple rSLSS counts &  50 &  13 &  6 &  4 &  13 &  14\\
%row no:14
 \  Density in Counts per gap binwidth &  8.0 &  8.4 &  6.0 &  4.0 &  13.0 &  8.3\\
%row no:15
 Loner rSLSS counts &  63 &  9 &  21 &  2 &  19 &  12\\
%row no:16
 Density in Counts per gap binwidth &  10.1 &  5.8 &  21.0 &  2.0 &  19.0 &  7.1\\
%row no:17
 \midrule
 rSLSS counts &  113 &  22 &  27 &  6 &  32 &  26\\
%row no:18
 Density in Counts per gap binwidth &  18.1 &  14.2 &  27.0 &  6.0 &  32.0 &  15.3\\
%row no:19
 pSLSS counts &  41 &  8 &  5 &  9 &  9 &  10\\
%row no:20
 Density in Counts per gap binwidth &  6.6 &  5.2 &  5.0 &  9.0 &  9.0 &  5.9\\
%row no:21
  \midrule
rSLBS counts &  27 &  5 &  7 &  5 &  6 &  4\\
%row no:22
 Density in Counts per gap binwidth &  4.3 &  3.2 &  7.0 &  5.0 &  6.0 &  2.4\\
%row no:23
 pSLBS counts &  7 &  0 &  0 &  3 &  3 &  1\\
%row no:24
 Density in Counts per gap binwidth &  1.1 &  0.0 &  0.0 &  3.0 &  3.0 &  0.6\\
%row no:25
 \midrule
 LSG counts &  123 &  18 &  39 &  46 &  11 &  9\\
%row no:26
 Density in Counts per gap binwidth &  19.7 &  11.6 &  39.0 &  46.0 &  11.0 &  5.3\\

\bottomrule
\end{tabular*}
\begin{tablenotes}%%[341pt]
\item Source: exoplanets.org, \citet{han14}, downloaded 2016 January 31. %Example for table source text.
% \item[1] Example for a first table footnote. %\item[2] Ex 2nd table footnote.
\end{tablenotes}
\end{table*}
\end{center}
%Earlier ver of tbl at:
  % ~/mg/awrite/distributions/submitgapreal/submitAstNach/tex_submitAstNach/hist_tex_submitAstNach/tbl_abbreviatedSelections.tex

%Resume here, include editing Fig. 3

% 2018endInto2019: Sec 2.1
\subsection{One large selection has a gap not present in other selections}\label{ssec_gapInOneSel}

We compare the distribution of counts per log period of rSLSS objects with
the distribution of other objects in Figure~\ref{fig_AllOthGap}
to show % that where we will show
(Section~\ref{sec_othrSngPile}) 
that other selections have a single pileup, the rSLSS population has a bimodal pileup separated into two peaks by a gap covering the period range from 493.7 to 923.8 day periods, which is 0.272 in log period or 0.16 of the ROI. We find the gap-width to be a useful unit to use to compare the density per log period of the peaks, since the ROI is 6.24 gap-widths wide. The mean density of the 113
%133 mistake
rSLSS objects is 18.1 objects per gap-width, so the six objects in the gap give it a density of only 0.33 of this mean. 
There are 49 (58) rSLSS objects in the ROI with periods shorter (longer) than the gap, a period range from 100 to 493.7 (923.8 to 5000) days that in log period covers 2.55 (2.70) gap-widths, which we call the short- (long-) periods side, [SPS (LPS)].
%      -Peaks have highest density next to gap which fall off to tails away from gap

Counts of how many objects are in the main parts of the
two peaks are given in Table~\ref{tab_CntDnsRgn}, where
we delineate both of the two regions at both sides of the gap into
one gap-width peaks, leaving tails of 1.55 and 1.70 gap-widths on the SPS and LPS sides, respectively, % in Table~\ref{tab_CntDnsRgn}, %T1
where we show the counts and densities of rSLSS objects 
per log period of the peaks, SPP and LPP, and tails, SPT and LPT, respectively.
The peaks have significantly higher densities
of rSLSS objects than any other region:
adjusting for the length of the tails, the tail versus peak density ratio is only 0.53 and 0.48 on the SPS and LPS
respectively.   \\

%(Metallicity Vs Period figure; Orig 2nd fig here:)

%Prev had here: Only Sunlike Objects, rSLSS&pSLSS Counts Vs Period figure; Orig 3nd fig here:

%\begin{figure*}[t]
%\centerline{\includegraphics[width=342pt,height=9pc,draft]{empty}}
%\caption{This is the sample figure caption.\label{fig1}}
%\end{figure*}

%~/mg/awrite/distributions/submitgapreal/submitAstNach/AN_LaTeX_template-1510016212000/tstltx/zzRmvdPcs_tpt.tex
%   Example for bibliography citations, downloaded from exoplanets.org, cite~\citep{han14}.

%\subsection{Example for second level head}

\subsection{Density of objects per log period in the two peaks versus in the gap} 
%in between them 
%\subsection{Numbers of objects in the two peaks and numbers of objects in the gap }
                                              %bm\_missingHowMany\tab \tab \par
   %Move to sssec_calcLikelyGap, Ln 744, Lkhd gp ran appear in obs if phys no gap
   %  Need check how use the density; missing actually giving density here
We quantify how deep the gap that is in between two peaks in the distribution of the periods of rSLSS objects by considering the density of periods per log period
in each of the peak and the gap regions.
%planets of stars of metal-rich sunlike single-stars (rSLSS), where we call the set of parameters associated with each planet  ``rSLSS objects''. 
This enables evaluation of how many objects are missing compared to if there were no gap by considering what the density is of objects per log period of the peaks on each side of the gap. 
%Didn't Rmv To Unlikely Section  -Moved back
%The mean density changes depending on how wide of a range from the gap we calculate the density, so we initially consider two ranges that are easy to discuss:\par
%Rmvd To Unlikely Section  -Start
%
%1.) A range of one in log period, and \par
%
%2.)\ a ``three-gap-width'' range going from one shallow gap width of 0.272 below to one width above the shallow gap, for a range of 0.816 in log period (three times 0.272).  %TDL  %  0.816344351
%Rmvd To Unlikely Section  -End

%Missing comparing the densities

\begin{figure}[t]
  \includegraphics[width=0.48\textwidth]{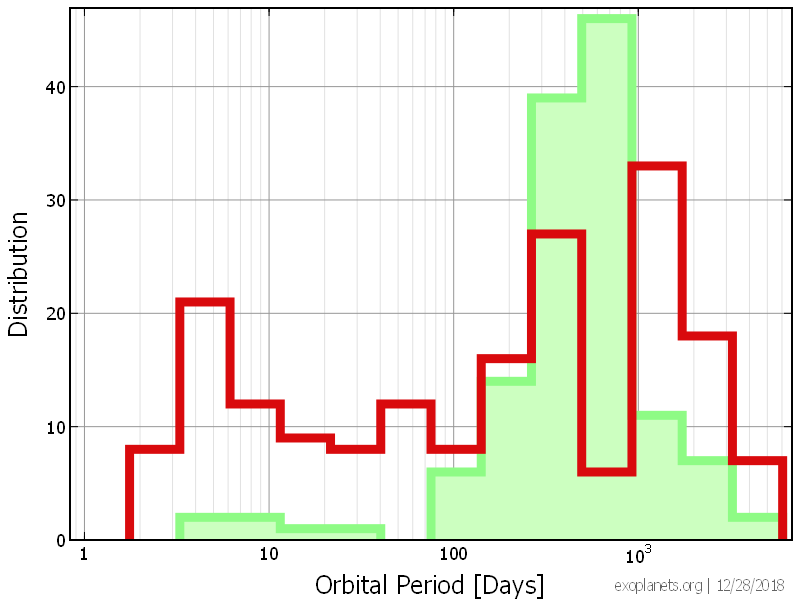}  %{media/ ...}
%	\centerline{\includegraphics[width=78mm,height=9pc,draft]{media/image1.png}}%.png
  \caption{The distribution of objects by log period of the rSLSS objects 
% (\textbf      
(red, unfilled) %rmvBold 2019-07   not of left parenthesis
%  \caption{The distribution of objects \textbf{by log period} of the rSLSS objects (\textbf{red, unfilled}) %rmvBold 2019-07  two
% }  rmvBold 2019-07 
    compared with those of the other objects (green, filled).  %rmvBold 2019-07 
%(\textbf{green, filled}).  %rmvBold 2019-07 
    When wide bins the width of the full (shallow) gap are
    used, the gap region, with six objects with periods between 493.7 days to 923.8 days, %rmvBold 2019-07  %rmvBold 2019-07
    still has a clearly lower density %\textbf{
of objects per log period
% }  rmvBold 2019-07  
of rSLSS objects. %rmvBold 2019-07
    \label{fig_rSlssXclVper}}
\end{figure}

In order to better show the shallow gap, in Figure~\ref{fig_rSlssXclVper}
we use the full width of the
shallow gap to bin the counts of periods of rSLSS objects 
%is demonstrated in Figure~\ref{fig_rSlssXclVper},
%where the counts of the rSLSS objects are binned with the full width of the gap
(red), 
which are compared with the counts of the ``others'' objects 
(green, filled).  %(green\textbf{, filled}). %rmvBold
Though this scale is too coarse to show the completely empty deep gap region,
it illustrates how an exceptionally low density region of planet periods per 
log period is bracketed by the two highest density regions.
%Treating the full gap as one wide bin of low density, as shown .\\

%2019-03: Mv 2.3 Lik of LoDensRgns towards end, first merge with 5.3 
%Mvd: \subsection{Likelihoods of low density regions} 

% (Should table be moved here along with density sections?)

\subsection{Peaks have highest density closest to gap } %\label{sec3}
%\section{Peaks have highest density next to gap }\label{sec3}
  %which fall off to tails away from gap

%\subsection{xx}

Even at the resolution of one-quarter of the gap, as shown in 
Figure~\ref{fig_AllOthGap}, the densities fall-off from their highest values next to the gaps. 
The quarter-gap-width bin in the SPP next to the gap, with
periods from 422.1 to 493.7 days, has 10 objects, or 2.2 times the mean density of the
ROI of 4.5 objects per bin of one-quarter the gap-width, followed with 7, 5, and 5
objects, comprising the 27 objects within one gap-width of the short period side of the
gap.
The quarter-gap-width bin in the LPP next to the gap, with periods from 923.8 to 1080.5
days, contains 13 objects, or 2.9 times the mean density of 4.5 objects per bin of this width,
followed by bins of 8, 8, and 3 objects comprising the 32 objects within one gap-width of
the log period side of the gap.

%Metallicity Vs Period figure; Now 2019-07 2nd fig Orig 2nd fig here:
\begin{figure}[t]   %  [Fe/H]* v Per
  \includegraphics[width=0.48\textwidth]{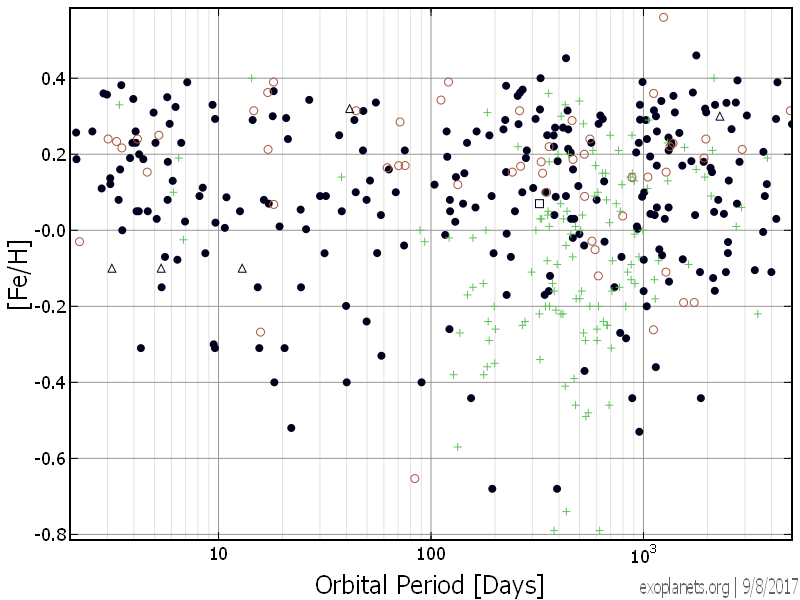} %{media/ ...} note 2<->3
%	\centerline{\includegraphics[width=78mm,height=9pc,draft]{media/image1.png}}%.png
	\caption{Plotting objects' metallicity versus period shows 
	   %rmvBold 2019-07   \textbf{
	that there exists an
	area that has a paucity of SLSS objects (filled black circles), 
	but where the density of LSG objects
	(green crosses) is not lower than elsewhere.
	 There are no SLSS objects with [Fe/H] greater than -0.03 in 
	 the range of periods from 653.22 to 923.8 days.
	 The presence of six such objects in periods from 493.7 days to 923.8 days
	 gives a much lower density than the adjacent areas.
	 %Table 1
	 In contrast, of the 41 SLSS objects with lower [Fe/H] within the ROI, 
	 nine are within these 
	 period range.
	 % has a gap in the SLSS objects (filled black circles) starts with [Fe/H] greater than -0.03,
	 %where the same period range  but with [Fe/H] less than -0.03 has a significant number of SLSS objects. 
	 %This range in is filled with LSG objects (green crosses), 
	 %and 
	 This area is also not empty of BS objects (red unfilled circles).
	    %  }  rmvBold above 2019-07
	 \label{fig_Fe_v_Per}}
\end{figure} 

%rSLSS& xcl Counts Vs Period figure;  %In first AstNach submission was Fig 5

%2019-03: Mvd from here  2.5 Deep part unlikely  
%Mvd:\subsection{Deep part of gap with zero objects unlikely in random distribution}\label{ssecDeep}

\subsection{Patterns consistent between radial velocity searches}\label{ssecObsvr} % between observing groups
% See ~/mg/awrite/distributions/sepfeats/gapfeats/observerChecking/tbls_observerCounting.doc  
% ~/mg/awrite/distributions/submitgapreal/submitAstNach/tex_submitAstNach/zTableCompose.tex

To consider whether the behavior of observers might have produced the
features found here, a comparison of the results from different 
groups of astronomers might have produced different patterns.
For example, if a group of observers conducted their RV search programs
for only a limited number of years and then quit, then 
their data could have a maximum period, 
with a cutoff beyond they would not report any more systems.
This apparent cutoff could show up in the data despite there being no such
cutoff in the distribution of periods of real planets.
This might account for the dropoff in periods beyond 493.7 days.
It is a little more difficult to say that an observing group might have
resumed its observations such that a sudden rise in the density of objects
is produced.
We studied the results from different observers whom we divided into two
groups, and found %\textbf{
that a peak-gap-peak pattern
% }  rmvBold 2019-07  %\textbf{similar} rmvBold 2019-07
appeared %\textbf{
separately % }% }  rmvBold 2019-07  
in both sets of data.   %rmvBold 2019-07
There may have been a different falloff of counts density at periods longer
than 1000 days, as might be expected if there was more of a falloff in
observing intensity of one group versus another.
For different groups, blind to the presence of these features and blind
to the other groups data, to produce data evidencing this peak-gap-peak 
pattern makes the chance that this pattern results from some behavior
of the observing very unlikely.

% stillDo 2019jun05        -- stillDo: Different observing groups see pks+Gap
An analysis of which radial velocity groups found how many objects in each of the peak, gap, and peak regions shows that the primary planet-finding groups %all 
maintained their observations for long enough to continuously find planets with periods 
longer than the long period pileup.
There certainly was no mass stoppage of observing to produce a false dropoff
in periods around 493.7 days,
nor was there any mass restart of observing that could have caused the long
period pileup.
%from shorter than the short period pileup, through the gap boundary periods of 493.7 through 923.8 days, and continuing past the long period pileup.

% resume 2018(?) Oct 31
Counting the numbers of which observing group discovered planets in the period
ranges from inner peak through the gap and into the outer peak
show no changes in the consistency of observers finding new planets.
The fraction of planets found among rSLSS objects by each of the 
%two or three
major groups that found planets remains similar from the short period peak
to the long period peak.

The largest two groups finding planets started with the first discoveries
of exoplanets.
We combine counts of the planet discoveries of the smaller groups with the 
two largest groups and show the sums of planets found in period range in
Table~\ref{tab_ObsvrCnts},
where the ``European'' group was started in 
Switzerland by Mayor and Queloz, ~\citep{may95},
and is dominated by groups on the European continent
except for the UK, 
which results from the ``Anglo-Australian'' collaboration we
combine with the planets found by the ``American'' group 
started by Marcy and Butler, ~\citep{mar96},
along with
results from a group from Japan and another U.S. group.
Though there are only six rSLSS objects in the gap, these
are also divided between the groups.

Both groups have found planets in all populations of periods well past
1000 days.
We find that all of the main groups who found planets with periods in
%264 days
the first peak, up to 493.7 days, continued finding planets through the
gap region, past the start of the deep gap at 923.8 days, and at
least up to one gap width longer in log period, up to periods of 1729 days.

We also find that these same groups were finding SLBS objects
and pSLSS objects, though there are fewer of these than rSLSS objects.
We find that as far as can be detemined with the smaller number
of these objects in the peaks, that the same groups continued to find planets
throughout the period regions of the peak-gap-peak features.
The  SLBS and pSLSS populations do, however, have more objects in the
gap region, and we see by the counts per group
that these observers found objects in all these period regions
at fractional rates similar to the fractions of total planets found
by each group.

%	263.85	493.70	923.80
%	493.70	923.80	1728.59
%	0.2721	0.2721	0.2721
%rSLSS	27	6	32
%pSLSS	5	9	9
%rSLBS	7	5	6
%pSLBS	0	3	3

%Count of observer groups table             Table~\ref{tab_ObsvrCnts}
\begin{center}
\begin{table*}[t]%
\caption{Counts of planet discoveries by observing groups.\label{tab_ObsvrCnts}}
\centering
\begin{tabular*}{500pt}{@{\extracolsep\fill}lccD{.}{.}{3}@{\extracolsep\fill}}
  \toprule
%&\multicolumn{2}{@{}c@{}}{\textbf{Spanned heading\tnote{1}}} & \multicolumn{2}{@{}c@{}}{\textbf{Spanned heading\tnote{2}}} \\\cmidrule{2-3}\cmidrule{4-5}
%\textbf{col1 head} & \textbf{col2 head}  & \textbf{col3 head}  & \multicolumn{1}{@{}l@{}}{\textbf{col4 head}}  & \textbf{col5 head} & \textbf{col6 head} & \textbf{col7 head}  \\ %& \textbf{col6 head}  \\
  %\midrule
  % From ~/mg/awrite/distributions/sepfeats/gapfeats/observerChecking/
  %   countObsPerRegion.xls sheet AuthTbl
%row no:1
 \textbf{Regions}  &  \textbf{SP Peak} &  \textbf{Gap} &  \textbf{LP Peak} \\
%row no:2
\midrule
 Start of period, days &  263.8 &  493.7 &  923.8 \\
%row no:3
 End of period, days  &  493.7 &  923.8 &  1728.6 \\
\midrule
 Counts total all groups  &   &   &   \\
%\midrule
 rSLSS               &  27 &  6 &  32  \\
 pSLSS               &   5 &  9 &   9  \\
 all SLBS            &   7 &  8 &   9  \\
% rSLBS              &   7 &  5 &   6  \\
% pSLBS              &   0 &  3 &   3  \\
\midrule
 Counts by European and associated groups  &   &   &   \\
 rSLSS               &  11 &  4 &  13  \\
 pSLSS               &   2 &  6 &   4  \\
 all SLBS            &   4 &  3 &   2  \\

\midrule
 Counts by U.S. and associated groups  &   &   &   \\
 rSLSS               &  15 &  2 &  19  \\
 pSLSS               &   2 &  3 &   4  \\
 all SLBS            &   3 &  2 &   4  \\

\bottomrule
\end{tabular*}
\end{table*}
\end{center}

%\section{
%\section{
\section{Planet-mass limited ``loner'' selection has decreasing tail to peak ratio}\label{sec_Loner}
%}\label{sec4}
%}\label{sec4}

Two more selections that further reduce the relative number of counts 
in the gap also enhance the peaks by producing 
a distribution of proportionately fewer 
objects in the tail regions away from the gap. 
We show that the ratios of the density of the tail counts divided 
by the density of the peak counts decrease for both the SPS and LPS. 
Using the densities in Table~\ref{tab_CntDnsRgn}, in the rSLSS selection of 113 objects, the ratio of the densities of the tail regions to the gap-width ``peak'' regions on the SP (LP) side is 0.53 (0.48).

\subsection{``Loner'' and Jupiter-mass selections}\label{ssec_loneMp}

When the rSLSS selection is divided into the 50 objects in multiple-planet and 63 objects in single planet systems 
%(rmSLSS or 
(``multiples'' and %rlSLSS or 
``loners''), % selections), 
the two distributions are strikingly different. 
Four of the six objects in the gap are among the multiples. %multiple-planet objects.
The distribution of multiples is more uniform, having smaller or no peaks
% if at all 
(Table~\ref{tab_CntDnsRgn}).

In the selection of 63 loner rSLSS objects, there are two objects that remain 
in the gap, which is a slightly lower relative density.
%  ???  What lower density
%% A inner ref to Methods that isnt unlikelihood.(Section~\ref{sec_Unlikely}) or  ~\ref{ssec_Unlikely}.
% (Methods). 
More significantly, both have $m$sin$i$ among the low and high extremes of the $m$sin$i$ distribution of rlSLSS objects. 
The features here are primarily features of the 56 objects of planet mass 
%not too much different from        %less or greater than
similar to Jupiter's. 

\begin{figure}[t]
  \includegraphics[width=0.48\textwidth]{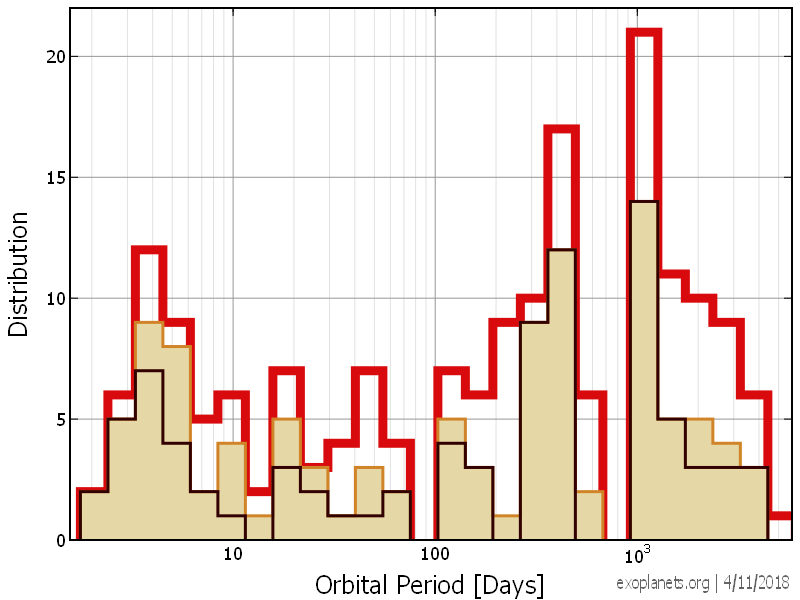} %{image5.png} %{media/ ...}
%	\centerline{\includegraphics[width=78mm,height=9pc,draft]{media/image1.png}}%.png
  \caption{The two peaks become more distinct in the loner rSLSS selection (orange), and even more distinct when a cut to intermediate $m$sin$i$ (black) of 0.3 < $m$ sin $i$ < 9.0 $M_J$ is performed. The rSLSS counts are shown in red. 
%  Quarter gap-width bins are used at top, 
  Half-gap width bins are used since these selections have closer to half as many objects as in the rSLSS selection, which gives numbers per bin that are similar to Figures~\ref{fig_AllOthGap} and \ref{fig_FeVsPer}. %at bottom. 
  \label{fig_rlJ}}
\end{figure}

Finally, it is expected that planet distributions depend on planet mass, and indeed, the only two objects remaining in the gap region are in the low and high extremes of the $m$sin$i$ distribution. One is slightly under 0.3 $M_J$, which is one of the two objects lowest in $m$sin$i$, and the other is nearly 9.5 $M_J$, one of the five objects highest in $m$sin$i$. 
This makes it possible to take cuts of a minimum $m$sin$i$ of 0.30 $M_J$ and a maximum $m$sin$i$ of 9.0 $M_J$ leaving the 56 middle-mass ``rich lone Jupiters,'' or ``rlJ'' objects.%, 
%``Jupiter'' $m$sin$i$ loner rSLSS objects (Jml-rSLSS objects, or ``metal-rich Jupiter-like loners'', rJLL) 
%of which histograms are shown in orange in Figure~\ref{fig_rlJ}.
%This suggests that the gap feature is associated with the 56 of the 63 loner objects with masses from 0.3 to 9.0 $M_J$. 
%Since the $m$sin$i$ of this region surrounds Jupiter's mass, we call these objects the ``rich lone Jupiters'' 
%(the Jupiter-mass-loner rSLSS), 
%or ``rlJ'' selection for short, since they are like Jupiter in planet mass (and the other SLSS selections), but are lone metal-rich objects. 
The distribution of the 56 rlJ objects are are shown in orange in Figure~\ref{fig_rlJ}.
 
Limiting our selection to these 56 rlJ objects with $m$sin$i$ in between the $m$sin$i$ of these two objects reveals that these ``middle-mass'' objects 
are even more concentrated in the two peaks   compared to %versus 
their tails (Figure~\ref{fig_rlJ} and Table~\ref{tab_CntDnsRgn}). 
%Wrongly printing as 2

\subsection{Relatively more objects in the two peaks in   ``Loner'' and Jupiter-mass selections}\label{ssec_PeaksDistinct} % Rmvd "the''
As we make the cuts going from the rSLSS to the loner selection, and then to the rlJ selection, each step removes more multiple objects from both tails than from the peaks. 
The result is that in the loner selection, the two peaks and gap become more distinct with each step. 
Using the ratio of the densities of the tail regions away from the gap to the peak regions near the gap shows how the two peaks on both sides of the gap become more dominant among the loner rSLSS and rlJ selections.

%"Rewrite"    SC 
Although these further two cuts were performed blind to the rest of the distribution,
they not only reduced the number of objects in the gap to zero, but 
reduced the number of objects away from the peak much more than in the peak.
This can be quantified by how these cuts reduced the tail to peak density ratio.

We use values from Table~\ref{tab_CntDnsRgn} to calculate that %Table 1 % (Methods) that
the tail to peak density ratios for the SPS and LPS respectively go from 0.53 and 0.48 for the rSLSS selection to 0.28 and 0.37 for the loner rSLSS selection, and then to 0.22 and 0.28 for the rlJ selection. 
%Given that 
%These selections were performed for how they reduced the objects in the gap to zero, but were blind to the tail to peak ratio. 
% ??  It appears that the populations in the tails and gap are separate from those in the peaks.    ??Populations question 2019-06-05

The two peaks are the primary features for these middle-mass single-planet metal-rich single-star objects, suggesting that for this population of objects, planet formation most generously produces planets in the period bands of these two peaks.

%    resume here (along with observing groups) 2019-06-05
\subsubsection{Mean densities including of the stricter selections}   %\par
The mean densities per gap width of the 
rSLSS selections, loner rSLSS, multiples rSLSS, and rlJ selections 
are the number of objects in each selection divided by 6.24 gap widths of the ROI. 
%rSLSS selections, rlSLSS, rmSLSS and rlJ

So the mean densities are 18.01 objects for the 113 objects in the 
rSLSS selection, 10.1 for the 63 objects in the loner rSLSS selection, 
8.0 for the 50 objects in the multiples rSLSS selection, and 
9.0 for the 56 objects in the rlJ selection.

Going from the rSLSS selection to the loner rSLSS selection 
reduces the relative density from 0.33 +- 0.14 to 0.20 +-0.14 
of the mean densities, 
while the relative density of the rmSLSS selection increases 
to 0.50 +- 0.25.

\section{Other populations have only one single pileup in region of the double-peaked pileup}\label{sec_othrSngPile}
%\section{Double-peaked pileup of major population where others have single pileup}\label{sec_othrSngPile}
%; No gap in the three other groups           12-27

The sum of the other selections shows that the same region is part of the peak of the other populations (Figure~\ref{fig_AllOthGap}, green filled-in bins). There is no dip in the LSG selection or in the pSLSS selection below [Fe/H] of -0.07. This region is also high in counts in the SLBS selection, though the counts are too low to consider smaller regions such as the deep part.  We look at counts in the gap region of the combined grouping of the 198 ``Other'' objects in the three other selections: the LSG, SLBS, and pSLSS selections. 

We show the metallicity distribution by period of all objects in Figure~\ref{fig_Fe_v_Per}. The black filled circles for the SLSS selection show that there is a gap region for metallicities above a boundary that actually is
closer to [Fe/H] of -0.03, though for simplicity we have used [Fe/H] of zero
in most of the presentation here.
%boundary that is a little below zero. 
The planets in ``other'' selections are those hosted by stars of low surface gravity (LSG, green crosses), stars that are sunlike (in surface gravity) but have at least a binary stellar companion (SLBS, red unfilled circles), or are sunlike single stars poorer in metallicity than the sun (pSLSS). (Squares denote two SLSS objects outside of ``sunlike'' effective temperatures, % $T_{eff}$, 
of 4500 < $T_{eff}$ < 6500 K.)

There are 63 ``other'' objects in the region of the gap in rSLSS objects,
%\textbf{
where % } % }  rmvBold 2019-07  
we sum values from the pSLSS, BS, and LSG populations %rmvBold 2019-07
(as shown in Table~\ref{tab_CntDnsRgn}).
%(Not SI? If SI, Formerly Table 1 of Supplementary Info or Methods, then tab2). 
Of these three other selections, when considered separately 
%% A inner ref to Methods that isnt unlikelihood.(Section~\ref{ssec_Unlikely}).
%(Methods), 
only the BS selection has a % any significant 
possible sign of a gap. 
So the rSLSS gap region is part of the peak in density for other regions, as there are nearly twice as many ``other'' objects as the ROI mean density, but only one third the mean density of rSLSS objects.

We will also use the ``deep part'' of the gap where there are %  ``Deep Part''
zero rSLSS objects to demonstrate how unlikely it is for there to be a region where there are no rSLSS objects, because it is especially straightforward to calculate the unlikelihood of having a region of objects every one of which is outside the rSLSS selection.
This deep part is the longward 0.55 fraction of the gap or 0.089 of the ROI, consisting of the region from the periods of 653.2 to 923.8 days.
If a region this wide were populated at the mean density, it would have 0.089 of the objects, or 17.5 ``other'' and 10.0 rSLSS objects, but there are 33 ``other'' objects 
%% A inner ref to Methods that isnt unlikelihood.(Section~\ref{ssec_Unlikely}).
%(Methods)
and 0 rSLSS objects. Since the deep part has close to twice (1.9) the mean number of ``other'' objects as the mean, it is again part of the peak of the other three selections.

The presence of so many objects in the other populations where there are few or none in the rSLSS population is evidence against the gap being an observational effect: %There
there could not be an observational failure to find rSLSS objects but not other objects in just one period range.
Though %\textbf{
these % } % }  rmvBold 2019-07 
data might be subject to selection effects due to having been %rmvBold 2019-07
collected by many different observers, it is extremely unlikely for any observational effect to reduce only the rSLSS population but not the other populations.

%Fig 4 2019-04-17
%4th fig here (or 5th if SuppInf fig put before this but either way still the two files  image4 and  image5):

%2019-03: Mvd from here the orig 4.1 Gap unlikely in ONLY one selection
%Mvd:\subsection{Gap unlikely to randomly appear in observations of only one selection}\label{ssec_Unlikely} %fix-tag :

%SI  start Supplementary Information
%\section{Review (from supplementary)}\label{sec_review} %\label{ssec_}
%\section{Unlikelihoods}\label{sec_unlikelihood} 
%\sub
\section{Unlikelihood demonstrated from more than one perspective}\label{sec_unlikelihood}%\label{ssec_more1persp}
                        %p7-tag 2018-12-26/27
We show that a gap this wide that divides the peak of the pileup into two peaks
is unlikely to be produced at random from two perspectives:\\
(1) Considering only the distribution of rSLSS objects alone, and\\
(2) comparing the distribution of rSLSS objects alone with the distribution
of the other selections.

In each case, we can calculate the likelihood of the existence of only the more 
narrow region with zero objects (the ``deep gap''),
or the wider full region between the peaks that has a lower density
but still has a small number of objects.
We also calculate how likely it is for both to exist.
We are careful in our calculations to count allowing the positioning of these gaps
to occur anywhere within a region containing at least the inner part
of two peaks, to not artificially require that the random features match
the specific positioning of the gaps that we find.

This wide of a gap separating the peak of the pileup into two peaks is unlikely to be produced 
at random by first considering the likelihood of a gap in distribution, and second, considering the likelihood of a consecutive string of values to entirely come from a subset of our full dataset.

The gap between the peaks is unlikely to be random from two perspectives: It is unlikely that the rSLSS population has the shallow and deep gaps, and it is unlikely that there is a wide region with no gap in the populations outside of the rSLSS population such that all of the 33 objects in the region of the deep gap come from the other 60$\%$  of the total population.

We discuss two important ways of looking at the peaks and the gap:\\
1. Emphasizing the separation between two peaks that has a low density (six objects), or\\
2. Emphasizing the gap as a region where the distribution of planets
has so few objects that from the current data there are zero objects.\\

To model how the gap has the two subregions,
one of low density and one of zero objects,
we consider two models based on two simplications of the gap:
(1) The wide low density region encompassing the full shallow gap, and 
(2) the narrow but deeper region encompassing the deep gap.
The deep gap encompasses a little more than half of the long period
part of the shallow gap.
Note that the sharpest transition is the boundary going from the
long period edge of both gaps to the
longer period peak, 
marked by the object having a period of 923.8 days.\\

%Treating the full gap as one wide bin of low density, as shown in 
%Figure~\ref{fig_rSlssXclVper}.\\

%polishing 2019-03-27  Dbl-check for re-insertion
% Can first be simplified twice, into separately looking at the wide and shallow full gap and the narrow-but-deep part of the gap. This is a little overly simplified, so we then try looking at the likelihood of having the full gap that has a more narrow deep part.

%Old loc of rSLSS& xcl Counts Vs Period figure; XclSlss.png}  %{media/ ...}

We first consider only the objects in the rSLSS population as distributed by log period, and show that the log period distribution gap is unlikely, especially in between two peaks.

We follow by considering how there are 33 objects in the main other selections in the period range of the deep gap where there are zero objects in the rSLSS selections. Since the rSLSS selection contains nearly 40$\%$  of the total objects, this means that the number of rSLSS objects in the range should be closer to 22 objects. It can be seen in 
%\textbf{the histogram of the total counts, Figure~\ref{fig_AllOthGap}}, %rmvBold 2019-07 
the histogram of the total counts, Figure~\ref{fig_AllOthGap}, %rmvBold 2019-07 
that there is in fact a dip in the total counts in the range of the deep gap. We cite the presence of this dip to argue that the deep gap of rSLSS objects cannot be due to incorrect measurements of surface gravity leading to misclassification of objects in this period range.

%2019-03: Mvd former 2.3 Lik of LoDensRgns towards end, first merge with 5.3 
%Mvd: \subsection{Likelihoods of low density regions} 

\subsection{Likelihoods of a ``gap'' or region of low density} 
\label{ssec_LikLoDens}
%Stephen: Not sure why don't like "list"
We list in  Table~\ref{tab_likelihoods} the   %not this link of Tbl1 {tab_CntDnsRgn}    % Supp Info Table 1 the 
likelihoods calculated how often random distributions produced gaps by taking two views of the 
gap, and then considering them together:\par

\begin{enumerate}
	\item The more narrow ``deep part'' of the gap that contains zero objects, of periods from 653.22 to 923.8 d. \par
	\item The wider full gap that contains six objects, with periods from 493.7 up to 923.8 days.\par
%	\item The wider gap with six objects containing a deeper gap within.
\end{enumerate}\par

The true likelihood would be somewhere in between choosing a range near the peaks of the distribution, where the gaps occur, and in choosing most of the ROI from 100 to 5000 day periods. Choosing most of the ROI underestimates the likelihood by allowing the gap to appear in a lower density of counts per log period than the gap actually occurs in, as well 
by allowing a gap to occur either on the short or long period end where the gap would merely cause a shorter or longer pileup rather than splitting the gap in two. Choosing a gap in too narrow of a range falsely reduces the range in which it may occur. 
We choose two reasonable ranges and suggest that the true likelihood is in between the two calculated likelihoods.\par
%Stephen: Not sure why don't like last "choose"

%resume here 2019-06-06: Remove repeat definitions of deep gap
%2019-03: Mvd to here former  2.5 Deep part unlikely 
%Mvd:
\subsubsection{Deep part of gap with zero objects unlikely in random distribution}\label{sssec_Deep}

Because all six of the rSLSS objects in the gap are in the short period side
in less than half of the shallow gap, the longer part of the gap
with zero rSLSS objects,
which we call the ``deep part'' of the gap.
% Late definition of ``deep part''?
These six have periods above the ending of the SPP at the period of 493.7 days,
to the longest at 653.22 days.
The deep gap with zero objects thus goes from the period of 653.22 to the starting
period of the LPP at 923.8 days,
%We refer to the 
%on the short period side from the boundary of the SPP at a period of 493.7 to the period of 653.22 days. We find a ``deep part'' of the gap from this period of 653.22 days that extends to the start of the LPP at the period of 923.8 days that has zero rSLSS objects. Though the shorter period ``shallow part'' of the gap contains multiple-planet rSLSS objects (below), the deep gap does not contain any of these rSLSS objects either. This deep gap 
which covers a distance of 0.150 in log period space, 
or 0.089 of the ROI, and which is 0.55 of the width of the shallow gap.
If a region with 0.089 of the 
ROI had the mean density of rSLSS objects in the entire ROI including the tails, 
it would have 10.0 objects.
Both the full gap and the deep part represent a significant deficit relative to even a uniform distribution, but this gap structure is most significant relative to how the two pileups that are on both sides have their highest densities of counts per log period next to the gaps. More than half of the objects in both peaks are within one shallow gap width of 0.272 in log period of the shallow gap.  
Of the 49 objects on the short period side of the gap, 27 are within this 0.16 %{0.1602}
of the ROI, and of the 58 objects on the long period gap, 32 are within this distance that has only six objects in the gap.
This means that if the shallow gap had the same density as the mean density of the same width on each side, it should have closer to 29.5 rSLSS objects, not just six.

We show in Section~\ref{sssec_calcLikelyGap} %{ssec_Unlikely}  % sec_Unlikely was sec8 %Methods 
that the likelihood of these features occurring at random starts at only 1 in 5000 when only parts of these features are considered, such as having the deep gap in a uniform distribution, but the likelihood becomes even less when considering more aspects of the distribution.

\subsubsection{Likelihood of gap randomly appearing in observations if the physical distribution has no gap}
\label{sssec_calcLikelyGap}

%  \begin{adjustwidth}{0.25in}{0.0in}
%  {\fontsize{8pt}{9.6pt}\selectfont $\textbackslash$ opt$ \{ $ ed$ \} $ $ \{ $ [Ed: Use as example current work showing that] $ \} $ \par}\par
%  \end{adjustwidth}

To see how likely this gap would appear at random, we conducted Monte Carlo simulations of either uniform or peaked distributions to see how often
an equivalently significant gap appears \textit{anywhere} within selected regions
of interest (ROIs) encompassing the peak. 
Though the likelihood varies depending on how wide of an ROI is chosen,
we find that the gap in the rSLSS
%this outer-period metal rich sunlike 
population is extremely unlikely to be a random feature, at the level of less than one in $10^4$ random instances.

If modeling that the likelihood  of a random gap appearing in
a distribution that is actually physically characterized by a single pileup,
this gap would %\textbf{
occur %}%}  rmvBold 2019-07 
somewhere near the center of the pileup. %rmvBold 2019-07
This means that we do not allow the gap to occur too close to either
the short period or long period edge.
This means that we calculate the chance that this gap will occur
only in a selected smaller region, and thus count only the number
of objects within this selected region.
It is a matter of judgment to consider how large of a subregion to choose,
but we choose subregions that are still large relative to the full pileup.
Rather than trying to argue for on size of a subregion,
we perform two calculations of unlikelihood is performed within the following two ranges
that we argue bracket the true unlikelihood with one value too high and one too low.

%Didn't Rmv To Unlikely Section
%This density changes depending on how wide of a range from the gap we calculate the density, so we initially consider two ranges that are easy to discuss:\par

  1.) A range of one in log period, and \par

    2.)\ a ``three-gap-width'' range going from one shallow gap width of 0.272 below to one width above the shallow gap, for a range of 0.816 in log period (three times 0.272).   %  0.816344351

The first subregion
is wide, allowing the gap to be far from the center
such that the true unlikelihood is probably higher than the calculated
value for this region.
The 2nd is small, where it might be argued that the 2nd 
small region presupposes the existence of the gap:

%  Put the two evaluations, log of one and three-gap-width, here
  %Move from line 367
We summarize the evaluation by Monte Carlo of several cases in Table~\ref{tab_likelihoods}.
For assessing the likelihood of an equivalent gap appearing at random when supposing the actual underlying population has no gap, we have developed a Monte Carlo program to test likelihood of an equivalent gap occurring with the same number of data points, $N_{range}$.
We computed what the width of an equivalent gap would be for a range from 0 to 1 first by considering the fraction that the gap width is of the width of the outer pileup, and then second calculating what fraction of the period
region that contains the peak of the pileup
%of the points in a Gaussian distribution
does this gap represent.
We are then able to use how often a random distribution of $N_{range}$ values
in between zero and one have a gap as large as this fraction
to evaluate how likely it is to have a gap occur near the middle
of the peak of the pileup.
The Monte Carlo program obtains the likelihood of the gap by,

  1.) Creating vectors of $N_{range}$ random values in between 0 and 1,

  2.) Taking the difference in between each point, and
then finding the largest difference.

  3.) Seeing what fraction of random vectors has a difference as large as the gap.

%             $N_{range}$

We  compare what likelihood is obtained this way,
when we use
the density of the peaks outside the gap while using the counts 
per log period from the one-log-period range from 200 to 2000, and 300 to 3000 day periods, and find it makes little difference. From 200 to 2000 day periods, there are
$N_{range}$ = 79 objects. %in our selection,
While this is a density of 79 objects per log period for this selected
period range,
there is a higher density for the part of the range outside the gap,
which has a length of 1.0-0.272=0.728 log period and has 73 objects after subtracting the six in the gap.
The density for just that part 
%[Stephen crossed out:] of the range outside the gap 
is 73/0.728, or a much higher value of 100.3 objects per log period.
From 300 to 3000 day periods, there are 78 objects in our selection.
%for similar values of densities.
%To consider only the density of objects in the region outside the gap, we do not count the six objects in the gap,
Here there are 72 objects in just the region outside the gap.
Outside the gap, there is a density of 98.9 objects per log period 
for the range 200 to 2000 days. So using either range we obtain a density in the peaks of close to 100 objects per log period.

We find that even in the case of allowing
a gap of zero objects with a width of 0.089         %gap of 0.1505 
in log period to appear 
anywhere in a distribution of the 79 objects
randomly placed in the range of 1 in log period, from 300 to 3000 days, that 
the frequency of the deep part of the gap randomly appearing is less than 
one in 5000 random instances.
The occurrence of the full (shallow) gap with six objects in a region of 0.1505
is a little more than 1 in 2000 instances. 
For a region with as low density as the deep part of the gap to occur anywhere within a wider shallow gap
of these widths happens only once in in $4 \times 10^4$ times. 

However, because a range of one in log period includes some of the
tails as well as the peaks,
this underestimates the likelihood because by allowing the gap to
fall anywhere, it could fall to the sides, in which case the gap would not 
really be creating two peaks.
We seek to bracket the actual likelihood
by also considering a more
narrow range.
Since this may overestimate how unlikely this feature is,
we believe that the true likelihood is somewhere in between the likelihoods
found using these two ranges.
We now consider the likelihood 
that the gap occurs near the peak of the pileup,
so that we require that the gap occur in slightly less wide regions bounded by the peaks.
We find using Table~\ref{tab_CntDnsRgn} that there are 65 objects
which are bounded within 
one full gap width of the gap, with periods from 263 to 1729 days,
which is a log period distance of 0.816.  %(Gap is of width 0.27211)
The 65 objects include the six in the gap, 
and 59 in the peaks which
comprise most of the high density regions.
We then evaluate the likelihood of the gaps occurring among 65 objects.
The shallow gap fills one third of this ``three-gap-width'' range,
and the deep part of the gap which is 0.55 of the gap width, 
fills 0.18 of this range.   % 0.184382 from 0.5531448/3.
The number of times that the shallow gap 
and deep part
occur only once in numbers that approach or exceed one in $10^5$ instances, 
depending on how fully we match the peak of the pileup to the double peaks.
%Stephen says depends, not depending
 
These results show that these gaps are extremely unlikely to be due to random.

%Moved 2019-03-27
The calculations considering the likelihood of a deep gap within the full gap
do not quite account for how other configurations of inner and outer gaps
would also be called unlikely. We could consider that the shallow part
might have more objects but that the deep gap is wider, or that the shallow part would have fewer objects but the deep gap is more narrow. 
These alternatives must be added into a two-level evaluation, 
but it is hard to assign the width of the deep gap in such cases. 
We could say that the actual likelihood of any surprising gap is perhaps 
two or three times greater than the one we consider here, but we suggest that
even if we did perform this two-level evaluation, 
we are still talking about such a very unlikely random gap that at this point, 
the main point is that this gap is extremely unlikely to appear at random.\\

% Likely NEW section addressing Others

% (Should table be moved here along with density sections?)
%%%%%%%%%%%%%%%%%%%% Table No: 2 (former SuppInfoTbl_1) starts here %%%%%%%%%%%%%%%%%%%%
% SI-TBL-1 HERE

\begin{center}
\begin{table*}[t]%
\caption{Summary of  likelihoods bracketed between calculations using wide and narrow ranges that the gap might appear Numbers of objects, fractional span of the gap, and likelihoods. Above: Fraction of random distributions that show a gap as large as the deep gap, the shallow gap, or both together. Below: The inverse fraction, or ``one in how many'' random distributions does the gap occur. \label{tab_likelihoods}}
\centering
\begin{tabular*}{500pt}{@{\extracolsep\fill}lccD{.}{.}{3}ccc@{\extracolsep\fill}}
\toprule
% \textbf{Regions}  &  \textbf{ROI} &  \textbf{SP tail} &  \textbf{SPP} &  \textbf{Gap} &  \textbf{LPP} &  \textbf{LP tail}\\

%row no:1
\textbf{Range} & \textbf{Number of objects} & \textbf{Deep-Part} & & \textbf{Full Gap} &  & \textbf{Both-Together}\\

%row no:2
Log-one in period & 79 & 0.151 & 2.0 e-4 & 0.272 & 5.5 e-4 & 2.5 e-5 \\
%row no:3
Three $ \times $  Shallow Gap & 65 & 0.184 & 1.1 e-4 & 0.333 & 2.2 e-4 & 1.2 e-5 \\
% \hhline{~~~~~~~}
%row no:4
% \hhline{~~~~~~~}
%row no:5
\textbf{One-out-of:} &  &  &  &  &  &  \\
%  \hhline{~~~~~~~}
%row no:6
\textbf{Range} &  & \textbf{Deep-Gap} &  & \textbf{Full Gap} & & \textbf{Both-Together} \\
%row no:7
 Log-one in period &  79 & & 5263 & & 1852 &  40000 \\
%row no:8
 Three $ \times $  Shallow Gap &  65 & & 9091 & & 4545 &  85000 \\
%\hhline{~~~~~~~}

\bottomrule
\end{tabular*}
\begin{tablenotes}%%[341pt]
%             \item Source: Example for table source text.
%   \item[1] Example for a first table footnote.
%   \item[2] Example for a second table footnote.
\end{tablenotes}
\end{table*}
\end{center}
% (End of section under question: Should table be moved here along with density sections?)

%Herebad; Move up to talk of the others

\subsection{``Other'' three selections have single pileup with no gap}
\label{ssec_othersNoGap}
%\textbf{No gap in the three other groups}

We compare how many objects in each group are in the region with how many objects in each group would be expected to be in a region this wide if the distribution were uniform in the ROI.

We consider whether the sum of all ``other'' objects has any gap. 
The ``other'' selection consists of 198 objects, summed 
% (Main Table~\ref{tab1}) 
from the 123 LSG, 34 SLBS, and 41 pSLSS objects in the ROI
(Table~\ref{tab_CntDnsRgn}). 
The mean density of the gap-width region that is 0.27 in log period, %( )
which is 0.160 of the 1.70 log period width of the ROI, %) 
is then 31.7 objects (of all types) per gap-width.

The shallow gap has only six rSLSS objects, compared to a mean of 
18.1 rSLSS objects per 0.27 log period, yet the same period region
from 493.7 to 923.8 days contains 63 ``others'' 
%\uline{63 ``others''} 
compared to a mean density of 31.7 other objects
per gap-width bin of 0.160 of the ROI. 
These 63 ``others'' break down into the other three populations, followed by the mean number of objects in a bin this size, as follows:\par

46 LSG objects versus a mean of 19.7 LSG objects;
%%Put back in: 46 LSG objects versus a mean bin count of 19.7

%%\tab 
%46 LSG objects versus a mean bin count of 19.7 \par
%    %{\fontsize{7pt}{8.4pt} \selectfont  19.70objects, \par}

8 SLBS objects versus a mean of 5.5 SLBS objects; and %{\fontsize{7pt}{8.4pt}\selectfont $ \{ $ 5.446$ \} $  5.5 objects, \par}\par

9 pSLSS objects versus mean of 6.6 pSLSS objects. %{\fontsize{7pt}{8.4pt}\selectfont 6.567objects. \par}\par

(If we subdivide the SLBS selection in order to compare metal-rich objects that 
have stellar companions with our rSLSS selection hosted by single stars, %: 
the 8 SLBS objects are composed of 3 pSLBS versus a mean of 
%{\fontsize{7pt}{8.4pt}\selectfont 1.21  
1.2 pSLBS objects per bin, 
and 5 rSLBS versus a mean of 4.3 
%{\fontsize{8pt}{9.6pt}\selectfont  4{\fontsize{7pt}{8.4pt}\selectfont .324 
objects per bin.
So, though the numbers are too low to give strong statistics, 
it appears probable that neither the metal-rich or metal-poor selections of the SLBS
selection shows a sign of a gap.)

All three of the ``other'' populations have a higher count in this 
period range, meaning the region of the shallow gap in the rSLSS population 
corresponds to a region that in all three other populations
is part of the peaks of those populations.

\vspace{\baselineskip}
%  DEEP GAP
% ~/mg/awrite/distributions/sepfeats/gapfeats/counts+densities.xls
When we narrow our evaluation to the deep part of the gap 
where there are zero rSLSS objects, the 0.151 log period between
periods of 653.2 %653.220
to 923.8 days, 
or 0.089 of the ROI,
compared to a mean of 10.01 ``other'' objects 
 %\uline{    
there are 33 ``other'' objects in this region. %} 
This compares to a mean density of 17.5 
%{%\fontsize{7pt}{8.4pt}\selectfont  17.541 
``other'' objects in a mean width of 0.089 of the ROI. These 33 others similarly break down as follows:

25 LSG objects versus a mean of 10.9 LSG objects; \par
    %{\fontsize{7pt}{8.4pt}\selectfont  10.896\ objects,  \par}
2 SLBS objects versus a mean of 3.0 SLBS objects; and \par
    %{\fontsize{7pt}{8.4pt}\selectfont  3.012 objects \par}
6 pSLSS objects versus a mean of 3.6 pSLSS objects. \par
    %{\fontsize{7pt}{8.4pt}\selectfont  3.632 objects.\par}

The LSG and pSLSS population densities are still high in the region of
the deep gap. 
The the two SLBS objects are a statistically insignificant one count
below the mean of three such objects in this region.
(Both are rSLBS objects.)
Whether this represents a partial gap among the SLBS population 
requires more data.

%The two SLBS objects are summed from 0 pSLBS objects, versus a mean count 
% of 0.62, and 2 rSLBS objects versus a mean count of 2.4 
%{\fontsize{7pt}{8.4pt}\selectfont  2.392}  
%objects per gap-width. %Though perhaps low, more statistics are needed.) 

The distributions of the other 198 objects in the three populations in the ROI do not have any such deep gaps. The 123 LSG objects in the ROI are enough to say this selection has no sign of any gaps. Though the number of objects are not high enough among the 41 pSLSS and 34 SLBS objects to 
be strongly significant,
it appears that there is no region with similarly few objects
in the distribution of periods of the LSG population. \par

The counts of LSG and pSLSS objects are still high compared to their means in this region; 
the count of two objects in the rSLBS selection in the region of the deep gap could be low,
but is not zero. This shows that there is no sign of a gap in the LSG distribution.
%{\fontsize{6pt}{7.2pt}\selectfont \sout{or in the pSLSS} population\sout{s}, with the gap region being part of the peaks of LSG\sout{and pSLSS}  objects.\par}\par

There is also no sign of a gap in the pSLSS population, though we see from Figure~\ref{fig_Fe_v_Per}
%\textbf{Fig. 2} %Main fig 2
that this best applies to moving the boundary of ``metal poor'' to
be at an [Fe/H] of below -0.07. The deep part of the gap region in the rSLSS population is part of the peak of pSLSS objects, especially if the boundary is taken %lowered 
lower to be at an [Fe/H] of -0.07.\par

For the 34 objects where the star has a stellar companion, the ROI contains 27 rSLBS objects and seven pSLBS
For the deep part of the gap region to contain two out of all 27 rSLBS objects in the ROI, 
%(the 34 objects in the SLBS selection are comprised of 27 rSLBS and 7 pSLBS objects), 
compared to zero of 113 rSLSS objects in this same period region, 
this could be an indication that the presence of a 
stellar companion could cause planet periods to evolve to partially fill in the gap. %Also, t
The two counts still produce a lower density of rSLBS objects
in the gap region, 
which density is much lower than the neighboring regions
that correspond to the peaks.
Due to the statistically small numbers of these rSLBS objects,
however, the most that can be concluded is that %it is possible that 
the rSLBS population could have a partial gap in the region of the deep part of the rSLSS gap.

%Mvd from here the orig 4.1 Gap unlikely in ONLY one selection
%Mvd:
\subsubsection{Gap unlikely to randomly appear in observations of only one selection}\label{sssec_UnlikelyJustOne} %fix-tag :
%  Say? one selection but not the others?
% large part of parameter space, but not appearing at all in most other parameter spaces
% The second perspective to evaluate whether the gap is either observational or could be produced randomly is to consider how what fraction of all the objects are or are not planets of metal-rich stars inside versus outside the gap.
It is unlikely that the deep gap would occur in the rSLSS selection of 113 of the 313 total objects with periods past 100 days, which is 36\% of the exoplanet population in the main region, but not occur in the distribution of the remaining 200 objects, or 60\% of the objects in the database. In the region of periods from 653.2 to 923.8 days where there are zero rSLSS objects, this means that the 33 other objects appear in a consecutive series of not having any rSLSS objects. 
For each of the 33 objects, there should be a 36\% chance of the object 
being an rSLSS object.
So the chance of having 33 consecutive objects % in a row 
that are not rSLSS objects is 
the value calculated by, \\
\begin{center}
280(1 - 113/313)\textsuperscript{33}=1.1 $\times$ 10\textsuperscript{-4},
\end{center}
%280(1 - 113/313)33=1.1 x10-4,\\
which is less than one random distribution in 9000.

This shows that it would be extremely difficult for an observational effect to %separate out 
create such a gap. 
%according to which objects are hosted by sunlike stars that are also iron-rich. 
The fact that this ``excluded selection,'' composed of LSG, metal-poor and binary star objects, has so many objects with the same period region that the rSLSS selection has a gap 
argues against that an observational effect to be responsible. 
Such an observational effect would have to make objects of stars that are sunlike partially undetected and partially misclassified, while simultaneously not affecting detections of objects that are too low in temperature, too high in surface gravity, or are too low in metallicity, as well as only partially affecting planets of stars that have a binary companion. If such an observational effect exists, it is important to understand it. We submit that such an effect is extremely unlikely, leaving the explanation that the gap is a physical feature.\par

%cp-loc

%5th but perhaps will be earlier fig here, the SuppInf fig; due to two images in prev fig this is 6th image but named   image6):
\begin{figure}[t]
  \includegraphics[width=0.48\textwidth]{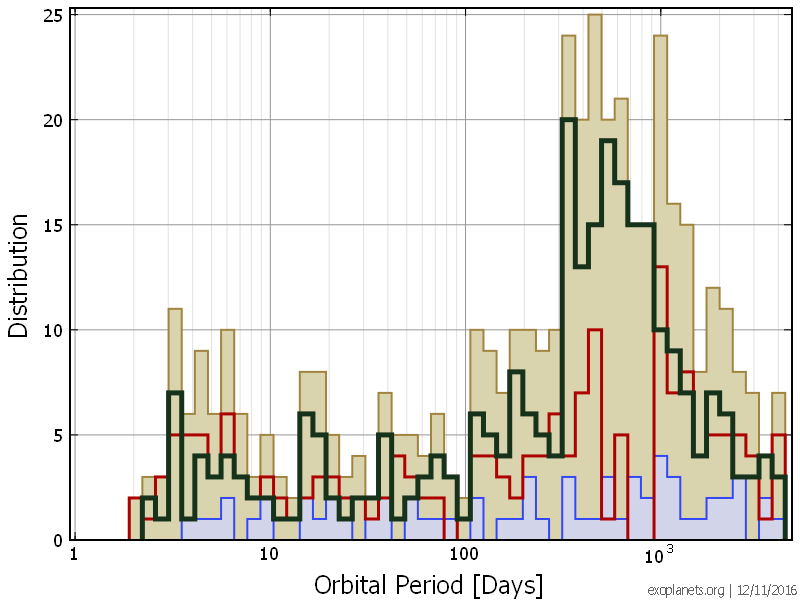}  %media/
%    \includegraphics[width=0.48\textwidth]{media/image6.png}
%	\centerline{\includegraphics[width=78mm,height=9pc,draft]{media/image1.png}}%.png
  \caption{ Counts of objects that are \textit{not} hosted by above solar metallicity sunlike stars, in thick green lines, compared to the total of all objects found by RV (light filled tan lines), metal-rich sunlike objects (red unfilled lines) and metal-poor sunlike objects (blue filled lines). \label{fig5}}
\end{figure}

% End SI Supplementary Information

\section{Discussion: What might cause the gaps and peaks?}\label{sec9}

The presence of these observational features challenges the current paradigm which maintains that planet formation is stochastic and cannot be regular.
It is especially difficult to reconcile such a sharp feature as the sudden rise in the number of objects which occurs at periods above 920 days.
It is difficult for a snowline trap to %\textbf{
play % }  rmvBold 2019-07 
such a role, since the snowline changes during 
evolution of the solar nebula. If the snowline does play a role, then it might require that there be a special time in this evolution that would trap planets at periods above 920 days.
It is even more difficult to suggest that planets either do not migrate from higher periods than 920 days down to lower periods, or that all planets that form at periods less than 920 days are migrated to periods longer than 920 days.

We also consider the model of~\citet{chr19}(hereafter CK19) 
%diffonly
who suggest that nebular material is trapped by density oscillations driven by the density of the material being described by the Lane-Emden equation oscillating around boundary conditions that the equation cannot satisfy.
The two peaks could be the two lowest period regions of such high density regions.
However, the CK19 hypothesis supposes a much higher degree of regularity of multiplanet systems than is generally accepted. 
The test of the CK19 hypothesis is to test it further on more multi-planet systems.

The regular features found here stand in contrast to the lack of regular features at longer periods of protoplanetary disks observed by the 
DSHARP project conducted by the ALMA observatory~\citep{and18,hua18}. 
%diffonly 
The resolution of ALMA is not high enough to resolve structures closer in than 10 AU, 
% (reference needed), 
and of course the DSHARP results are observations of disk material rather than planets, so the results here do not disagree with the DSHARP results. Still, this discrepancy implies that the inner later positioning of planets is more regular than the outer earlier positioning of nebular material. We encourage efforts to image protoplanetary disks 
%(PPDs)   %diffonly
at higher resolution.

These results show the importance of continuing to find more giant planets at periods of hundreds to thousands of days which have measurable parameters. The current data set is dominated by stars of mass not too different from solar mass, so it would help to find more planets of lower mass stars to better see how these features depend on stellar mass.

Planets found using the RV method are ideal in that the system parameters are generally well measured.

% Test~\citep{gon97,sant03,Fis05,udr03}.
It is expected that WFIRST~\citep{spe15}
%diffonly 
will find many planets within this region, but to evaluate these patterns it is important that the planet-hosting stars be recovered and more system parameters than just the planet and star masses be measured. It is most important to measure the period, stellar surface gravity, and the metallicity. Even without measuring the stellar surface gravity and metallicity, when the bulk planet set is plotted by period there exists an identifiable notch in the location of the gaps. So even if only the period is found, the WFIRST results may contribute to understanding these features.

\section{Summary}\label{summarySec}

%The planets of sunlike metal-rich objects, where 
The log period distribution of these objects has a bimodal pileup as opposed to the single pileup of iron-poor objects. 
In contrast, the distribution of periods of LSG and pSLSS objects 
beyond 100 days have single pileups. 
Though the number of BS objects is small, 
this group has a few objects within the gap,
so it is not clear whether the periods of BS objects have a 
partial double peak distribution or have a single pileup.

We show that looking at the distribution of periods of the rSLSS selection alone 
it is unlikely that a gap this wide would randomly show up in 113 objects. We also consider how unlikely it is 
within this range corresponding to 33 objects that there is a string of 33 objects all outside the rSLSS selection. 
This is not only unlikely to occur at random, but it is also unlikely that observational effects would somehow not find the objects in the rSLSS selection without decreasing the counts outside the rSLSS selection.
We also address the possibility of observational effects by comparing the distributions found by different groups of observers. Other than possible dropoffs at longer periods due to reduced observations, the planets found by
different groups follow the same general patterns of a valley up to perhaps
200 days with a pileup at longer periods.
We see that the distributions from different observer groups similarly show
this gap-peak-gap feature, demonstrating that it is unlikely that
the manner in which observations were taken falsely generated this pattern.

% We explore the structure

Features of the peaks that indicate underlying structure include
that the long period gap is most dense near the boundary with the gap
taken to be the period of 923.8 days. 
Though the gap is found in both single and multiple planet systems,
the two peaks are relatively even more prominent when only single-planet systems
are considered.
The only remaining two objects in the gap have one with a low mass
and one with a high mass, % outer regions of the mass range, 
suggesting that the full gap is a feature of planets with
mass not too much more or less than Jupiter's.
% Not one of the 41 objects in adjacent bins has a lower or higher mass
% 21 in the bin in the short period peak, 
% 20 in the bin in the long period peak, 

In contrast, there is only a single peak of LSG and pSLSS objects.
Among the rSLBS population, though there are objects within the gap,
there could be two separate peaks in the rSLBS selection.
More SLBS objects are needed to see the shape of the distribution
of SLBS objects.

% Composing in .../hist_tex_submitAstNach/twoPkGap_summaryPcs.tex
We find that the peak-gap-peak pattern is unlikely to occur either as
random or as a result of observations.
To see how likely these gaps are in just the rSLSS distribution,
we tested random distributions of the peak parts of the distribution,
consisting of more than half of the 113 objects
within the periods of 100 to 5000 days,
which is a log period span of 1.7. 
We found that even in a log period span of 1.0,
that it is unlikely to have a log period distance of 0.15
of the distribution not have any objects
at the level of one in many thousands.
It is similarly unlikely that a log period distance of 0.272
will have only six objects.
When considering all the populations of objects,
we find it unlikely that there will be a consecutive string of 33
objects without any rSLSS objects.

\section{Conclusions}\label{sec5}
%Uniform patterns.

That these patterns show up in the aggregate of %from 
so many different systems is evidence that planet formation and 
evolution must have much more uniformity than previously assumed. % \textbf{
Perhaps the biggest surprise, as gauged by the responses 
to when these results have been presented, is that planet evolution
is thought to be too chaotic to preserve such distinct features. %}  rmvBold 2019-07
While it is not surprising that there is an increase in counts 
corresponding to an important condensation temperature, the presence of such a deep gap at periods shortward of the main peak, with another shorter period pileup, represents a new major unexpected feature of the 
distribution of a very large fraction of exoplanetary systems. %\textbf{
How these features form, and how they remain preserved in
the cumulative distribution, are now important questions that must
be answered if we are to understand planet formation. %} %rmvBold 2019-07

%StartBigCut2 %\EndBigCut2

%\backmatter

\section*{Acknowledgments}

We are grateful for the support of Emily Chang.
This research has made use of the Exoplanet Orbit Database and the Exoplanet Data Explorer at exoplanets.org (Han et al. 2017).
We are grateful to Artie Hatzes, Dan Fabrycky, Guillermo Gonzalez, 
Edward Guinan, %\textbf{
Nader Haghighipour, Wen-Ping Chen, %}  rmvBold 2019-07 
Stephen Cooley, and Rosemary Mardling for helpful % \textbf{
responses and %}  rmvBold 2019-07} 
comments. % on the writing.

%This is acknowledgment text.%~\cite{Paivio1975}. Provide text here. 

%This work was performed under.% the auspices of the \fundingAgency{National Nuclear Security Administration} of the US Department of Energy at Los Alamos National Laboratory under Contract No. \fundingNumber{DE-AC52-06NA25396}.
 
\subsection*{Author contributions}
S.F. Taylor found the features, analyzed the data, performed the statistical analysis, and wrote the paper.

\subsection*{Financial disclosure}

None reported.

\subsection*{Conflict of interest}

The authors declare no potential conflict of interests.

\nocite{*}% Show all bib entries - both cited and uncited; comment this line to view only cited bib entries;
\bibliography{twoPkGap_aph}      %old {twoPkGap} {diff_twoPkGap} (nachImport}%

%\section*{Author Biography}
%(if applicable)
%
%\begin{biography}{\includegraphics[width=60pt,height=70pt,draft]{empty}}{\textbf{Author Name.} this is sample author biography text .}
%\end{biography}

\end{document}